# Plasmonic Waveguides to Enhance Quantum Electrodynamic Phenomena at the Nanoscale

Ying Li and Christos Argyropoulos, *Senior Member, IEEE*

*Abstract*—The emerging field of plasmonics can lead to enhanced light-matter interactions at extremely nanoscale regions. Plasmonic (metallic) devices promise to efficiently control both classical and quantum properties of light. Plasmonic waveguides are usually used to excite confined electromagnetic modes at the nanoscale that can strongly interact with matter. The analysis of these nanowaveguides exhibits similarities with their low frequency microwave counterparts. In this article, we review ways to study plasmonic nanostructures coupled to quantum optical emitters from a classical electromagnetic perspective. These quantum emitters are mainly used to generate single-photon quantum light that can be employed as a quantum bit or "qubit" in the envisioned quantum information technologies. We demonstrate different ways to enhance a diverse range of quantum electrodynamic phenomena based on plasmonic configurations by using the classical dyadic tensor Green's function formalism. More specifically, spontaneous emission and superradiance are analyzed by using the Green's function-based field quantization. The exciting new field of quantum plasmonics will lead to a plethora of novel optical devices for communications and computing applications operating in the quantum realm, such as efficient single-photon sources, quantum sensors, and compact on-chip nanophotonic circuits.

## I. INTRODUCTION

Light can couple to metal electrons along a metal-dielectric interface to form a surface wave. This wave is based on the collective electron oscillation and is called surface plasmon polariton (SPP) [1]–[3]. It is characterized by intense electromagnetic fields confined in a subwavelength region that decay quickly away from the interface. Due to their unique properties, SPPs have found a broad range of applications in various areas of science, including light harvesting, energy transfer, biochemical sensing, medical science, and high-resolution imaging [4]–[7]. Surface plasmon waves can also serve as an additional energy decay channel for quantum optical emitters located at their near field, leading to more efficient generation of quantum light or, equivalently, single-photon stream radiation [8]–[11]. They promise to open new routes towards the manipulation and boosting of several inherently weak quantum electrodynamic phenomena [12], [13].

Quantum electrodynamics is the study of quantized optical radiation and its statistical properties and interaction with materials [14]. It is mainly dedicated to the generation, manipulation, control, and entanglement of photons that are envisioned to serve as quantum bits or "qubits" in quantum information processes. Photons are ideal carriers of information because they are fast, robust, and capable of long-distance travel. They can also have different polarization states providing an additional degree of freedom in the efficient transfer of information and its computation. Integrated optical components operating with photons based on plasmonic (metallic) nanostructures are expected to have a compact footprint with nanometer dimensions combined with extremely low power consumption. The ultrafast and coherent nature of plasmonic-based light-matter interactions promises to overcome the quantum decoherence problem [15] that consists the major concern towards the practical application of several important quantum technologies.

The electrodynamics of dissipative (lossy) media, such as metals at optical frequencies, can be described by the widely used in electromagnetic engineering Green's tensor formalism which satisfies Maxwell's equations [16], [17]. The Green's function is a 3-by-3 dyadic tensor quantity that characterizes the impulse response of an electromagnetic system, where each column of the tensor is the induced electric field produced by an electric dipole polarized along the corresponding coordinate system axis. The quantization of the radiation field is based on the classical Green's function representation of the vector potential, identifying the external sources as noise that are usually associated with the loss of the dissipative medium [18]–[22].

In the quantum realm, spontaneous decay is generated by both vacuum fluctuations of the field and radiation reaction. This quantity can be drastically modified by shaping the material properties of the surrounding environment and the resulted phenomenon is called Purcell effect [23]–[25]. For example, plasmonic waveguides or nanoantennas can efficiently tailor the spontaneous emission properties of an emitter with respect to free space due to their ability to improve photon collection efficiencies [26]–[29]. The relevant quantity

This work was partially supported by the National Science Foundation (DMR-1709612) and the Nebraska Materials Research Science and Engineering Center (MRSEC) (grant no. DMR-1420645).

Ying Li is with the School of Physics and Optoelectronic Engineering, Nanjing University of Information Science and Technology, Nanjing 210044, China.

Christos Argyropoulos is with the Department of Electrical and Computer Engineering, University of Nebraska-Lincoln, Lincoln, NE, 68588, USA. E-mail: christos.argyropoulos@unl.edu

here is the local density of states (LDOS), a proportionality constant that characterizes the light-matter interaction strength and the modification in the resulting emission rates.

Even though many coherent light-matter interaction processes, such as spontaneous decay, are essentially non-classical, thus requiring a full quantum description, the media the quantum emitter is coupled to can be rigorously incorporated into the quantum formalism through the classical Green's function $\overline{\mathbf{G}}$ [24], [30]. Indeed, the entire information required to characterize several quantum electrodynamic processes can be encapsulated in the classical electromagnetic Green's function formalism [31]. More specifically, $\overline{\mathbf{G}}$ imaginary part can describe the LDOS [32], resulting in the spontaneous emission rate calculation [27], [33], while $\overline{\mathbf{G}}$ real part is capable to describe the photonic Lamb shift, a resonant frequency shift effect caused by the coupling of the emitter's bound electrons to vacuum modes [34], [35]. Furthermore, the two-point Green's function (also called mutual density of states) hold the information of the signal's propagation response from an emitting to a detection point [36]–[38].

All these quantum optical quantities are directly correspondent to the classical electromagnetic theory widely used by the electromagnetic engineering community [39]. Therefore, the emerging field of quantum plasmonics has attracted and will continue to attract considerable interest in the electromagnetics and quantum optics research communities [40]–[43]. The existing development of plasmonic nanostructures has been found to considerably boost many fundamental quantum electrodynamic phenomena mainly based on dipole-dipole interactions, such as van der Waals forces and vacuum friction [44], Förster energy transfer [19], [45], individual or collective spontaneous emission [46]–[48], and quantum information protocols like the realization of quantum phase gates [17], [49] and quantum entanglement [50]–[53].

## II. Dyadic Green's function

### A. Derivation of dyadic Green's function

To determine the dyadic Green's function of the electric field, we start with the wave equation in a homogeneous medium:

$$\nabla \times \nabla \times \mathbf{E}(\mathbf{r}) - k^2 \mathbf{E}(\mathbf{r}) = i\omega\mu_0\mu \mathbf{j}(\mathbf{r}), \quad (1)$$

where $\mathbf{j}(\mathbf{r})$ is an arbitrary current source distribution that can be viewed as a superposition of multiple point current sources. We replace the source term $\mathbf{j}(\mathbf{r})$ in Eq. (1) by the Dirac-delta function representing a point source $\delta(\mathbf{r} - \mathbf{r}')$ and define for each direction a corresponding Green function. For example, Eq. (1) becomes in the $x$- direction:

$$\nabla \times \nabla \times \overline{\mathbf{G}}_\mathbf{x}(\mathbf{r},\mathbf{r}') - k^2 \overline{\mathbf{G}}_\mathbf{x}(\mathbf{r},\mathbf{r}') = \delta(\mathbf{r} - \mathbf{r}')\mathbf{n_x}, \quad (2)$$

where $\mathbf{n_x}$ represents the unit vector along the $x$-direction and $\delta(\mathbf{r} - \mathbf{r}')$ is the Dirac-delta function representing a single point current source. As can be seen by Eq. (2), the Green's function is the induced field resulted from a delta function excitation. Similarly, we can formulate other two equations for a point source polarized along the remaining $y$-direction and $z$-direction. After accounting for all orientations, we can express the general electric dyadic Green's function as:

$$\nabla \times \nabla \times \overline{\mathbf{G}}(\mathbf{r},\mathbf{r}') - k^2 \overline{\mathbf{G}}(\mathbf{r},\mathbf{r}') = \overline{\mathbf{I}}\delta(\mathbf{r} - \mathbf{r}'), \quad (3)$$

where $\overline{\mathbf{I}}$ denotes the unit dyadic and $\overline{\mathbf{G}}$ is a dyadic Green's tensor whose $i$th column represents the electric field because of an arbitrary source polarized along the $i$th direction. It is worth mentioning that, in quantum optics, sometimes different constants are introduced on the right-hand side of Eq. (2) by convention, leading to different constant coefficients in the dyadic Green's function shown in Eq. (3). However, the electric field remains the same in both notations.

As shown in Fig. 1, once we know the dyadic Green's function $\overline{\mathbf{G}}$ of the electromagnetic system under study, we can find the electric field by the integration of the product of Green's function $\overline{\mathbf{G}}$ and source term $\mathbf{j}(\mathbf{r})$ over a volume $V$ and using the wave equation (1):

$$\mathbf{E}(\mathbf{r}) = i\omega\mu_0 \int_V \overline{\mathbf{G}}(\mathbf{r},\mathbf{r}')\mu(\mathbf{r}')\mathbf{j}(\mathbf{r}')dV'. \quad (4)$$

where $\mathbf{r}$ represents the position of the evaluated field point and $\mathbf{r}'$ designates the location of the point source (see Fig. 1). However, we know that the general solution of the inhomogeneous wave equation (1) consists of a homogeneous solution ($\mathbf{j}(\mathbf{r}) = 0$) and a specific inhomogeneous solution given by Eq. (4). Therefore, we need to add the homogeneous solution $\mathbf{E_0}$ in Eq. (4) and the resulted general electric field solution will have the final form:

$$\mathbf{E}(\mathbf{r}) = \mathbf{E_0} + i\omega\mu_0 \int_V \overline{\mathbf{G}}(\mathbf{r},\mathbf{r}')\mu(\mathbf{r}')\mathbf{j}(\mathbf{r}')dV', \quad (5)$$

while the corresponding magnetic field will be:

$$\mathbf{H}(\mathbf{r}) = \mathbf{H_0} + \int_V \left[\nabla \times \overline{\mathbf{G}}(\mathbf{r},\mathbf{r}')\right]\mathbf{j}(\mathbf{r}')dV', \quad (6)$$

where $\mathbf{E_0}$ and $\mathbf{H_0}$ are the electric and magnetic fields in the absence of the current $\mathbf{j}$. The above two equations are called volume-integral equations [24]. They are very important in electromagnetic field theory because they form the basis of various widely used theoretical and numerical modeling techniques, such as the method of moments and the coupled dipole method [1], [3]. Note that the following two relations are always valid for Green's functions applied to reciprocal systems:

$$\overline{\mathbf{G}}^*(\mathbf{r},\mathbf{r}',\omega) = \overline{\mathbf{G}}(\mathbf{r},\mathbf{r}',-\omega),$$
$$\overline{\mathbf{G}}(\mathbf{r}',\mathbf{r},\omega) = \overline{\mathbf{G}}(\mathbf{r},\mathbf{r}',\omega), \quad (7)$$

where $\overline{\mathbf{G}}(\mathbf{r},\mathbf{r}',\omega)$ represents the classical Green's function when the field propagates from a dipole source $\mathbf{r}'$ to $\mathbf{r}$.

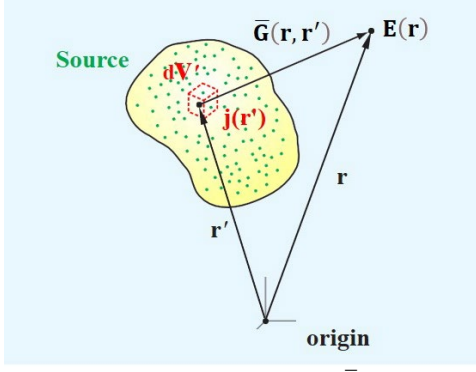

Fig. 1. Illustration of the dyadic Green's function $\overline{\mathbf{G}}(\mathbf{r}, \mathbf{r}')$, where the electric field $\mathbf{E}(\mathbf{r})$ in one arbitrary point can be determined through the integration of $\overline{\mathbf{G}}(\mathbf{r}, \mathbf{r}')$ and a distributed source. Reproduced from [24].

*B. Electric and magnetic fields computed by Green's function*

We compute first the electromagnetic fields induced by an electric dipole source embedded inside a homogeneous, local, linear, and isotropic medium. The current density $\mathbf{j}$ in the wave equation (1) can be regarded as an oscillating electric dipole located at the charge current center $\mathbf{r}_0$:

$$\mathbf{j}(\mathbf{r}) = -i\omega \mathbf{p} \delta(\mathbf{r} - \mathbf{r}_0), \quad (8)$$

where $\mathbf{p}$ is the electrical dipole moment of the dipole source. Therefore, after introducing the dipole current density of Eq. (8) into the volume-integral equations (5)-(6) and assuming that all the electromagnetic fields are produced from the electric dipole source (i.e., $\mathbf{E_0} = \mathbf{H_0} = 0$), we get the simplified relations:

$$\mathbf{E}(\mathbf{r}) = \omega^2 \mu \mu_0 \overline{\mathbf{G}}(\mathbf{r}, \mathbf{r}_0) \mathbf{p},$$
$$\mathbf{H}(\mathbf{r}) = -i\omega \left[ \nabla \times \overline{\mathbf{G}}(\mathbf{r}, \mathbf{r}_0) \right] \mathbf{p}. \quad (9)$$

Hence, the fields introduced by any arbitrarily polarized dipole source $\mathbf{p}$ placed at $\mathbf{r}_0$ can be determined by the dyadic Green function $\overline{\mathbf{G}}(\mathbf{r}, \mathbf{r}_0)$. For example, the Green function in free space (vacuum) can be derived by using Eq. (3) and is equal to:

$$\overline{\mathbf{G}}^0(\mathbf{r}, \mathbf{r}_0, \omega) = \left( \overline{\mathbf{I}} + \frac{1}{k^2} \nabla \nabla \right) \frac{e^{ikR}}{4\pi R}$$
$$= \frac{e^{ikR}}{4\pi R} \left[ \frac{(kR)^2 + ikR - 1}{(kR)^2} \overline{\mathbf{I}} + \frac{3 - 3ikR - (kR)^2}{(kR)^2} \frac{\mathbf{R} \otimes \mathbf{R}}{R^2} \right], \quad (10)$$

where $\mathbf{R} = \mathbf{r} - \mathbf{r}_0$ and $\mathbf{R} \otimes \mathbf{R}$ is the outer product of the distance vector $\mathbf{R}$. The imaginary part of the vacuum dyadic Green's function at the dipole source point $\mathbf{r} = \mathbf{r}_0$ is $\mathrm{Im}[\overline{\mathbf{G}}^0(\mathbf{r_0}, \mathbf{r_0}, \omega)] = k/6\pi$, representing purely radiative losses, since free space is lossless. The real part of the dyadic Green's function (i.e., $\mathrm{Re}[\overline{\mathbf{G}}^0(\mathbf{r_0}, \mathbf{r_0}, \omega)]$) is infinite at the same point, because of the divergent nature of the homogeneous Green's function in vacuum.

In the general case of inhomogeneous and lossy media, the dyadic Green's function, i.e. the solution of Eq. (3), can be represented as [30], [54]:

$$\overline{\mathbf{G}}(\mathbf{r}, \mathbf{r}_0, \omega) = \overline{\mathbf{G}}^{\mathrm{hom}}(\mathbf{r}, \mathbf{r}_0, \omega) + \overline{\mathbf{G}}^{\mathrm{sc}}(\mathbf{r}, \mathbf{r}_0, \omega), \quad (11)$$

where the first term $\overline{\mathbf{G}}^{\mathrm{hom}}(\mathbf{r}, \mathbf{r}_0, \omega)$ is the analytically known expression in homogeneous and infinite media (for example, free-space $\overline{\mathbf{G}}^0(\mathbf{r}, \mathbf{r}_0, \omega)$ with expression given by Eq. (10), while the second term $\overline{\mathbf{G}}^{\mathrm{sc}}(\mathbf{r}, \mathbf{r}_0, \omega)$ accounts the contribution from the radiative process of scattering due to the inhomogeneous scenarios, such as plasmonic waveguides, nanoparticles or other nanostructures [17], [30], [34].

III. QUANTUM PHENOMENA ANALYZED BY GREEN'S FUNCTION

The quantization scheme [18] of the dyadic Green function-based electric field can be employed to analyze a diverse range of quantum electrodynamic phenomena, such as spontaneous emission, collective spontaneous emission or superradiance, and other effects based on coherent dipole-dipole interactions of quantum dipole emitters coupled to any electromagnetic system. One of the remarkable properties of the dyadic Green's function is that it can be directly related to the LDOS, the spontaneous decay rate and the Lamb shift, as well as the spectrum of a quantum emitter observed at a detector. This scheme is widely used in the literature because it is semi-classical, simple to implement, and was found to accurately describe the response of quantum emitters when they are coupled to a dispersive and lossy environment. In quantum optics, the quantum emitters are always surrounded by an omnipresent fluctuating electromagnetic field, also known as vacuum state fluctuations (more details can be found in a relevant review paper [14]). This field is always there and fluctuates, even when the surrounding space is in its lowest energy state, called vacuum state, where no photons are present, and no light can be detected. These vacuum state fluctuations lead an emitter to decay "spontaneously" to a lower state, an effect also known as spontaneous emission [55]. The quantum emitter can be excited optically from the ground to the excited state leading to a spontaneous single-photon emission [56]. One of the most widely used electric field quantization approaches is based on the Green's functions Langevin local quantization [14], [57], which is extremely useful for calculating the lifetime and Purcell factor by adopting the Fermi's golden rule [24]. The quantum operator of the electromagnetic field in the presence of the plasmonic reservoir by using the Schrödinger equation approach is given by [13], [17]:

$$\hat{\mathbf{E}}(\mathbf{r}) = i\sqrt{\frac{\hbar}{\pi \varepsilon_0}} \int_0^\infty d\omega \frac{\omega^2}{c^2} \int \sqrt{\mathrm{Im}(\varepsilon(\mathbf{r}', \omega))} \overline{\mathbf{G}}(\mathbf{r}, \mathbf{r}', \omega) \hat{f}(\mathbf{r}', \omega) d\mathbf{r}', \quad (12)$$

where $\hat{f}(\mathbf{r}, \omega)$ is the bosonic field operator that plays the role of the local annihilation operator of the field excitation. Moreover, $\varepsilon(\mathbf{r}', \omega)$ is the complex permittivity of the surrounding space. Although the spontaneous emission process is in principle a non-classical procedure and requires the quantum operator description to compute its quantities, the coupling between emitter and medium can be rigorously characterized by the dyadic Green function formalism. Under the point-dipole approximation [58], which assumes that the quantum emitter can be modelled as a point dipole source, the local density of states (LDOS) of the electromagnetic modes is directly proportional to the imaginary part of the corresponding Green's function, and the resulted individual spontaneous decay rate can be computed by [24], [59]:

$$\gamma_{sp} = \frac{\pi \omega_0}{3\hbar \varepsilon_0} |\mathbf{p}|^2 \rho(\mathbf{r_0}, \omega_0), \tag{13}$$

$$\rho(\mathbf{r_0}, \omega_0) = \frac{6\omega_0}{\pi c^2} \left[ \mathbf{n_p} \cdot \mathrm{Im}\{\overline{\mathbf{G}}(\mathbf{r}_0, \mathbf{r}_0, \omega_0)\} \cdot \mathbf{n_p} \right], \tag{14}$$

where $\rho(\mathbf{r_0}, \omega_0)$ is the LDOS of the two-level system depending only on the position $\mathbf{r_0}$ of the dipole source, $\mathbf{n_p}$ is an unit vector directed to the electrical dipole moment $\mathbf{p}$ ($\mathbf{p} = |\mathbf{p}|\mathbf{n_p}$), $\hbar$ is the reduced Planck's constant, and $\omega_0$ is the atomic transition frequency of the quantum emitter. The spontaneous decay definition given by Eq. (13) is based on Fermi's golden rule which is valid in the weak coupling regime [12], [13], [24]. While the strong coupling analysis requires more sophisticated approaches involving the solution of time-dependent equations of motion [57], [60]. Note that the total spontaneous decay rate for plasmonic (lossy) systems needs to be divided into radiative $\gamma_r$ and nonradiative $\gamma_{nr}$ contributions [59]: $\gamma_{sp} = \gamma_r + \gamma_{nr}$. The radiative and nonradiative rates characterize the radiation and dissipation, respectively, processes of the generated photons by a quantum emitter. The ratio between the radiative decay $\gamma_r$ and the total decay $\gamma_r + \gamma_{nr}$ is defined by the quantum efficiency $QY$ of radiative decay (also known as quantum yield):

$$QY = \frac{\gamma_r}{\gamma_{sp}} = \frac{\gamma_r}{\gamma_{nr} + \gamma_r}. \tag{15}$$

Hence, the quantum efficiency provides a metric of the radiation performance of a plasmonic system combined with a quantum emitter, similar to the antenna efficiency metric widely used in microwave frequencies [39]. More details about how to enhance the spontaneous emission with plasmonic structures are provided later in Section V.A of this review paper. Note that by inserting the dyadic Green's function in free space with expression (10) into Eq. (14), the well-known value of an emitter's LDOS in free space can be derived to be [24]:

$$\rho_0 = \frac{\omega_0^2}{\pi^2 c^3}, \tag{16}$$

and the spontaneous decay rate in free space is equal to:

$$\gamma_{sp}^0 = \frac{\omega_0^3 |\mathbf{p}|^2}{3\pi \varepsilon_0 \hbar c^3}. \tag{17}$$

Apart from LDOS, which can be used to compute the spontaneous decay rate of a single isolated quantum emitter, the non-local density of states (NLDOS) can also be calculated to evaluate the resultant density of states caused by interference phenomena and coherent dipole-dipole interactions between two emitters. The total collective spontaneous emission rate can be computed by the NLDOS as [36], [50]:

$$\gamma_{ij} = \left(2\omega_0^2 / \varepsilon_0 \hbar c^2\right) \mathrm{Im}[\mathbf{p}_i^* \cdot \overline{\mathbf{G}}(\mathbf{r}_i, \mathbf{r}_j, \omega_0) \cdot \mathbf{p}_j]. \tag{18}$$

For emitter $i$, $\mathbf{p}_i^*$ represents the complex conjugate of the electrical dipole moment. Equation (18) is a more general version of Eq. (13), since it provides a more convenient way to calculate the emission decay caused by both self-interactions ($\gamma_{ii}$) and mutual-interactions ($\gamma_{ij}$), where $\gamma_{ii}$ is also known as the spontaneous decay rate given before by expression (13). As an example, $\gamma_{12}$ represents the collective contributions to the emission decay originating from the interference of emitter 1 placed at position $r_1$ with emitter 2 placed at position $r_2$. These mutual interactions are used to compute the effect of collective spontaneous emission or superradiance, as it will be demonstrated later in Section V.B of this review paper.

On a relevant context, the discovery [61] and explanation [62] of the vacuum Lamb shift lies at the foundation of modern quantum electrodynamics. The Lamb shift $g_{ii}$ is manifested as a resonant frequency shift in the emitter's atomic transition frequency caused by the coupling of its bound electrons to the surrounding vacuum state fluctuations [34]. The Lamb shift induced by a homogeneous medium can be incorporated in the definition of the emitter's resonant frequency. It is calculated by using the scattering or real part of the Green's function given by the formula [34]:

$$g_{ij} = \left(\omega_0^2 / \varepsilon_0 \hbar c^2\right) \mathrm{Re}[\mathbf{p}_i^* \cdot \overline{\mathbf{G}}(\mathbf{r}_i, \mathbf{r}_j, \omega_0) \cdot \mathbf{p}_j]. \tag{19}$$

Hence, $g_{ii}$ represents the photonic Lamb shift due to the self-interaction of each quantum emitter (qubit) with the environment. Note that this coherent quantity is proportional to the real part of the dyadic Green's function and is different compared to the decay rates $\gamma$ presented before, which are mainly based on incoherent processes. The coherent dipole-dipole interactions are characterized by $g_{ij}$, a property analogous to the real part of the Green function The $g_{ij}$ coefficient represents the coupling between emitters placed in spatial points $r_j$ and $r_i$ and can compute other interesting coherent emitter interaction processes, such as the Förster resonance energy transfer and quantum entanglement [53]. The dyadic Green function $\overline{\mathbf{G}}(\mathbf{r}_i, \mathbf{r}_j, \omega_0)$ satisfies the classical Maxwell's equations, as was shown in Section II [24]. Therefore, both Eqs. (18) and (19) can be calculated either analytically [17] or numerically [30] through solving classical Maxwell's equations. From the computational electromagnetics viewpoint, various numerical methods exist to calculate the Green's function in the case of inhomogeneous environment. Differential-equation-based methods [63], such as finite difference [64] and finite element [65], can be implemented but require to precisely discretize the scatterers and background, resulting in a large computational domain [66]. Integral-equation-based methods [67], such as the surface integral equation technique, also known as boundary element method, only discretize the scatterers and, thus, do not require any additional domain truncation or absorbing boundary conditions to achieve more accurate results [68]. In the case of numerical modeling, the derived by full-wave simulations electric fields can be used to calculate both real and imaginary parts of the dyadic Green's function by using Eqs. (9).

## IV. QUANTUM EMITTERS

As structures become smaller and smaller and reach nanoscale dimensions, their quantum behavior becomes apparent. The discrete nature of atomic states dominates in resonant light-matter interactions at the nanoscale. In atoms, molecules, and nanoparticles, these resonant interactions occur when the photon's energy matches the energy difference between their electronic energy levels. Due to the resonant character of these interactions, various quantum emitters can be approximated as effective two-level atomic systems where only two electric energy levels are considered whose energy difference is near the radiating photon energy. Two-level emitters can be used as photonic qubits, which are the main building blocks of quantum optical technologies. They consist an ideal system for the exploration of quantum optical phenomena in solid state physics [24], especially when used as indistinguishable single photon sources due to their on-demand and high-rate single photon generation capabilities [69]. The indistinguishability of these solid-state emitters is largely limited by dephasing that can be mitigated by using an optical cavity in both photonic and plasmonic systems [15], [70], [71].

Mainly, three types of quantum emitters exist: i) fluorescent organic dye molecules, ii) semiconductor quantum dots, and iii) impurity centers (a.k.a., color centers) in wide-bandgap semiconductors, such as diamond. In the following, we present a brief overview of these different quantum emitter types.

### A. Fluorescent organic dye molecules

The lowest-energy electronic transition of an organic dye molecule occurs between the highest occupied and lowest unoccupied excited state molecular orbitals [72]. The radiative relaxation between these orbitals is called *fluorescence* and is one of the most widely used radiative process. Fluorescent molecules have two decay contributions to their spontaneous emission process: i) radiative (i.e., fluorescence) and ii) non-radiative (through quenching or dissipation to heat) decay. Figure 2 schematically demonstrates the different processes and energy-level diagrams of an organic molecule, where the solid and dotted arrows represent radiative and non-radiative processes, respectively [72]. Since the electron is usually excited to a vibrational state, the fluorescence emission is redshifted to a lower energy level than the excitation (i.e., Stokes shift). In the case of organic dye molecules, the radiative decay dominates compared to nonradiative effects (such as quenching or dissipation), and the radiative lifetime is typically on the order of nanoseconds [24]. Aromatic or conjugated organic molecules exhibit particularly efficient fluorescence and are usually named as *fluorophores* (or *dye molecules*).

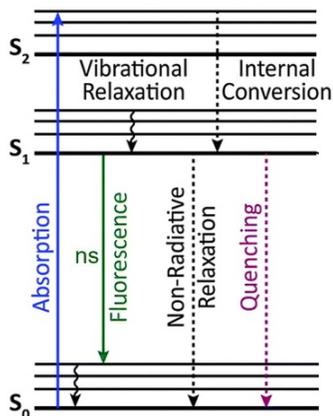

Fig. 2. Energy level diagram showing fluorescence and other nonradiative relaxation processes in an organic dye molecule. Reproduced from [72].

### B. Semiconductor quantum dots

Semiconductor crystallites are more widely known as quantum dots (QDs). They have nanoscale sizes with a radius approximately equal to their excitons' Bohr radius, which is the distance of an electron-hole pair. Their radius can be on the order of few to 10 nm mainly due to the small effective mass of their electrons and holes. Hence, notable quantum confinement can be achieved in QDs at length scales 10-100 times larger compared to other typical molecules [24]. As shown in Fig. 3, QDs have many remarkable quantum characteristics that distinguish them from bulk semiconductors, such as size and surface effects, strong quantum confinement, and macroscopic quantum tunneling. These effects lead to a splitting in energy levels from continuous (bulk semiconductor) to discrete (QDs), as depicted in Fig. 3. In the limit of extremely small dimensions, electron and hole pairs in QDs can be represented by a particle-in-a-box intuitive model, leading to a discrete energy level that can shift into higher energies as the box dimensions become smaller. Therefore, the energy level gap can easily be tuned by adjusting the crystal's size, which, subsequently, leads to the efficient control of the absorption and luminescence spectra. For instance, in the case of CdSe/ZnS QDs, a continuous change in the emission color is achieved just by decreasing the QD size [73]. In addition, the quantum efficiency of their radiative decay is rather high, mainly due to the strong confinement of both electrons and holes in a nanometer volume, making QDs extremely interesting for optoelectronic applications [74].

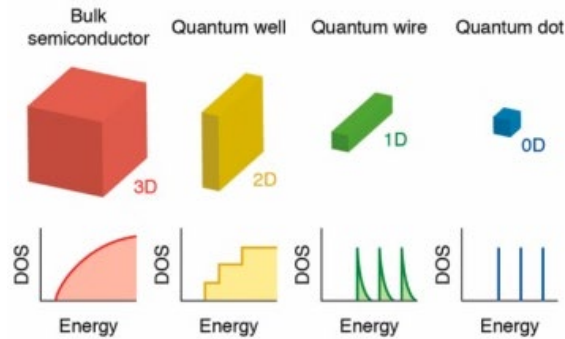

Fig. 3. Energy levels of several semiconductor nanostructures with different dimensionalities. Reproduced from [75].

### C. Color centers

The third class of quantum emitters is the most recently investigated and is composed of fluorescent defect centers in wide-bandgap semiconductors. For instance, diamond offers the largest bandgap (5.5 eV) of all known materials and hosts more than a hundred known luminescent defect centers. As shown in Fig. 4(a), the prominent impurity-related defect center in diamond is the nitrogen-vacancy (NV) center, which consists of a substitutional nitrogen atom (N) and a vacancy (V) at a nearest-neighbor lattice position. Such NV centers can form naturally during diamond growth or artificially using a variety of implantation and annealing techniques [24]. The NV center contains two unbounded electrons originating from the substitutional nitrogen. In addition, there are three more unbounded carbon electrons, where two of them form a quasi-bond and one remains unbounded. Therefore, the NV center can efficiently trap an additional electron, which turns into the negative NV⁻ center as opposed to the neutral NV⁰ center. Figure 4(b) shows a simplistic energy-level diagram of the NV⁻ center, where the NV⁻ center is treated as a three-level electronic system having a ground $|g\rangle$, excited $|e\rangle$, and intermediate $|s\rangle$ state. The main transition ($|g\rangle$ to $|e\rangle$) exhibits a zero phonon line at 637 nm combined with vibrational side bands in the range of 630-800 nm [76]. The radiative lifetime is around 13 ns for NV centers in bulk diamond and around 25 ns for NV centers in nanodiamond due to their different refractive index [24]. Very recently, color centers have also been explored in ultrathin 2D materials with wide band gaps, such as hexagonal boron nitride (hBN) [77]. These 2D materials consist another exciting new quantum emitter platform that can be easily embedded in photonic structures [78].

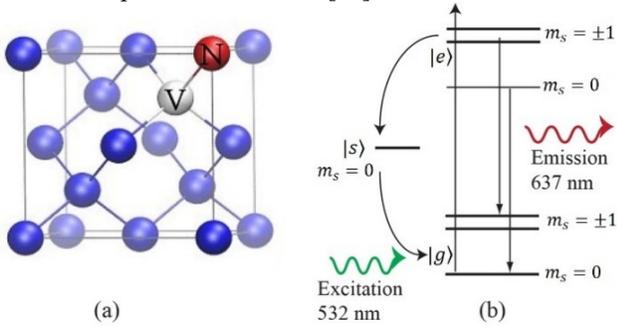

Fig. 4. NV center in diamond. (a) Schematic of the NV⁻ center in a diamond lattice showing the substitutional nitrogen atom (N) and the vacancy (V). (b) Energy level diagram of NV⁻ center, where $m_s = 0, \pm1$ represents the number of allowable spin states. Reproduced from [79].

## V. PLASMONIC QUANTUM ELECTRODYNAMICS

As discussed in the previous Section III, the dyadic Green's function, widely used to solve classical electromagnetic problems, can also be utilized to model various quantum electrodynamic effects. In the following, we present how several quantum electromagnetics phenomena will be boosted mainly by plasmonic nanowaveguides and their analysis by making use of the Green-function formalism presented before.

### A. Spontaneous emission rate enhancement

The major drawbacks of the presented in Section IV quantum light emitters are their relatively long radiative lifetimes (around 10 ns) and non-directional radiation. These lead to slow light generation response accompanied by weak fluorescence power. Hence, the intrinsic optical properties of quantum emitters cannot satisfy several demands of nanophotonic quantum optical devices [80], such as ultrafast light-emitting diodes, plasmonic nanolasers, and single-photon sources. In 1946, Purcell demonstrated that the spontaneous emission decay of a quantum source is not an intrinsic property but can be largely modified when the emitter is located inside a cavity due to the inhomogeneity induced by its interaction with the surrounding environment [23]. Therefore, many photonic resonance cavity systems, such as nanocavities [26], photonic crystals [81], nanostars [82], [83], plasmonic waveguides [34] and nanoantennas [69], [84], [85], were successfully used to enhance the spontaneous decay rate with the goal to achieve ultrafast operation comparable to high-speed optical networks. As an example, Fig. 5(a) shows the fluorescence emission of a single molecule as a function of its distance to a gold nanoparticle [58]. Metallic nanoparticles or films usually exploit the large LDOS at their SPP resonance frequency to achieve strong emission enhancement. However, this resonant enhancement is in turn restricted by the narrow bandwidth and high ohmic losses at the plasmonic resonance frequency mainly leading to detrimental nonradiative decay effects. As shown in Fig. 5(a), by varying the distance between the molecule and plasmonic nanoparticle, a continuous transition from fluorescence enhancement to fluorescence quenching was observed due to competing effects between the increased excitation rate and the nonradiative energy loss.

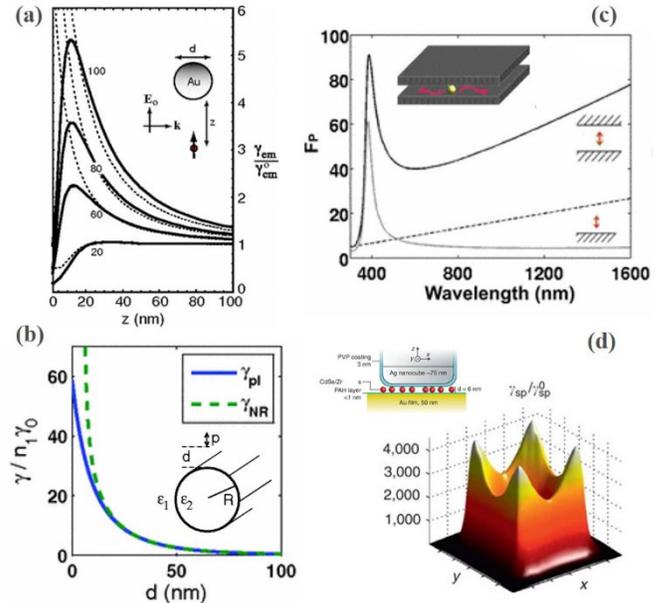

Fig. 5. (a) Total spontaneous emission (dashed) and radiative rate (solid) enhancement as a function of emitter-spherical gold nanoparticle separation distance. (b) Plasmon radiation rate (solid) and total nonradiative rate (dashed) of an emitter placed on top of a silver nanowire waveguide with different separation distance. (c) Spontaneous emission enhancement versus the wavelength for a metal-dielectric-metal parallel plate waveguide (dark solid line) and a metal-dielectric planar interface (gray solid line). (d) Map of computed spontaneous decay enhancement for a gap plasmon nanocavity loaded with an emitter. Figures reproduced from: a [58]; b [86]; c [87]; and d [69].

In addition to the many studies of simple metallic nanoparticles, there have been several recent investigations on spontaneous emission in plasmonic waveguide and nanocavity systems. For instance, it was demonstrated that the spontaneous emission of a single emitter in the vicinity of a plasmonic nanowire can be enhanced at the surface plasmon resonance, enabling various quantum optical applications [17], [86]–[89]. Figure 5(b) shows the plasmonic radiative enhancement and total nonradiative rate for an emitter located on the top of a plasmonic cylindrical nanowire. The total spontaneous decay rate consists of radiative and nonradiative decay rates. Note that 60-fold or even higher enhancement in the emitter spontaneous emission was achieved as the distance between the emitter and nanowire was decreased [86]. The emission enhancement of a point-like dipolar emitter was also investigated in a variety of plasmonic waveguide systems with arbitrary shapes [86]. To achieve an even tighter confinement of the plasmon waveguide modes, metal-dielectric-metal parallel plate waveguide structures were also explored [87]. As shown in Fig. 5(c), compared to a single metal-dielectric planar interface, there was stronger broadband nonresonant enhancement for an emitter placed inside the parallel plate plasmonic waveguide. Finally, Fig. 5(d) demonstrates that the spontaneous emission enhancement can approach very high values (4000) in the case of emitters coupled to the ultrathin metallic gap between a silver nanocube and a gold substrate [69].

Next, we consider an alternative design of an orthogonal plasmonic waveguide array operating near the cutoff frequency. One unit cell of the plasmonic waveguide geometry is shown in Figs. 6(a)-(b). The plasmonic grating consists of periodic slits with dimensions: width $w = 200$ nm, height $t = 40$ nm, and length $l = 500$ nm. These periodic nanowaveguides are made of silver (Ag) and are filled with a dielectric material (for example, glass). Each unit cell period is selected to be equal to $a = b = 400$nm. Quantum emitters, such as QDs or fluorescence dyes, are embedded inside the waveguide channels. The waveguide width is designed to achieve the cutoff frequency of the dominant quasi-TE$_{10}$ mode along the nanochannel at $f \approx 295 THz$, where the real part of the wave number becomes zero ($Re(\beta) = 0$) [90]. At this cutoff frequency ($f \approx 295 THz$), the plasmonic grating can be replaced by an effective epsilon-near-zero (ENZ) material, as depicted in Fig. 6(c). Moreover, due to the inverse relation between guided wave number $\beta$ and impedance $Z$, the corresponding characteristic impedance $Z$ is also very large at the ENZ cutoff, resulting in anomalous impedance matching that produces counterintuitive transmission accompanied by almost infinite phase velocity and uniform field distributions inside the nanochannels [91], [92]. Figure 6(d) shows a uniformly enhanced electric field in the orthogonal nanochannel's $yz$-cut at the ENZ frequency ($f \approx 295 THz$), indicating that largely enhanced coherent interactions between different emitters can be achieved when placed inside this ENZ plasmonic system.

Assuming operation in the weak coupling (Markov approximation) regime, a single quantum emitter embedded inside the channel is regarded as a two-level dipole source satisfying the electric point-dipole approximation [58]. The emission frequency of the emitter is chosen to be almost equal to the ENZ cutoff frequency. The orientation of its dipole moment is along $z$-axis in order to guarantee the maximum coupling with the waveguide. Full-wave 3D simulations are used to calculate the Green's function (Eq. (9)) and, as a result, the spontaneous decay rate and LDOS given by Eqs. (13) and (14), respectively. Based on these calculations, Fig. 6(e) plots the computed normalized spontaneous emission rate $\gamma_{sp}/\gamma_{sp}^0$ distribution at the ENZ resonance by changing the location of the emitter on a 4×50 grid inside the waveguide channel, where $\gamma_{sp}^0$ corresponds to the free-space spontaneous emission given before by Eq. (17). Note that the spontaneous emission can reach high values up to 200 compared to an emitter placed in free space and shows a uniform distribution.

However, as was mentioned before, in metallic systems, the total spontaneous decay rate needs to be divided into radiative and nonradiative contributions [59]: $\gamma_{sp} = \gamma_r + \gamma_{nr}$. The nonradiative decay rate $\gamma_{nr}$ can be evaluated with numerical full-wave simulation by integrating the absorbed power of the metallic parts [59]. The calculated total spontaneous decay rate and nonradiative rate result in the computation of the quantum efficiency $QY$, defined before by Eq. (15), which reflects the radiative emission efficiency, similar to the radiation efficiency of an RF antenna [39]. The development of future integrated quantum photonic circuitry will require efficient coupling between quantum emitters and nanophotonic materials. The $QY$ distribution is computed and shown in Fig. 6(f) and is found to have very high and constant values close to 0.7. Such large and uniform spontaneous emission rate enhancement and quantum efficiency makes the ENZ plasmonic waveguides excellent candidates for boosting the efficiency of various quantum electrodynamic effects [34], [48], [53]. Finally, it is worth mentioning that the aforementioned spontaneous emission effect for several quantum emitters is obtained in the weak coupling regime, where the currently used Markov approximation is valid. In the weak coupling regime, losses dominate, and the emission spectrum is directly related to the field confinement inside the plasmonic waveguide. On the contrary, in the strong coupling regime, the coupling outperforms the losses and the Markov approximation of spontaneous emission is expected to break down [12]. In this case, a double peak emerges in the emission spectrum because of the strong emitter-resonator interference. Strong coupling based on quantum emitters is one of the current hot topics in quantum plasmonics [43], where researchers strive to achieve the strong coupling regime by using plasmonic resonators [93].

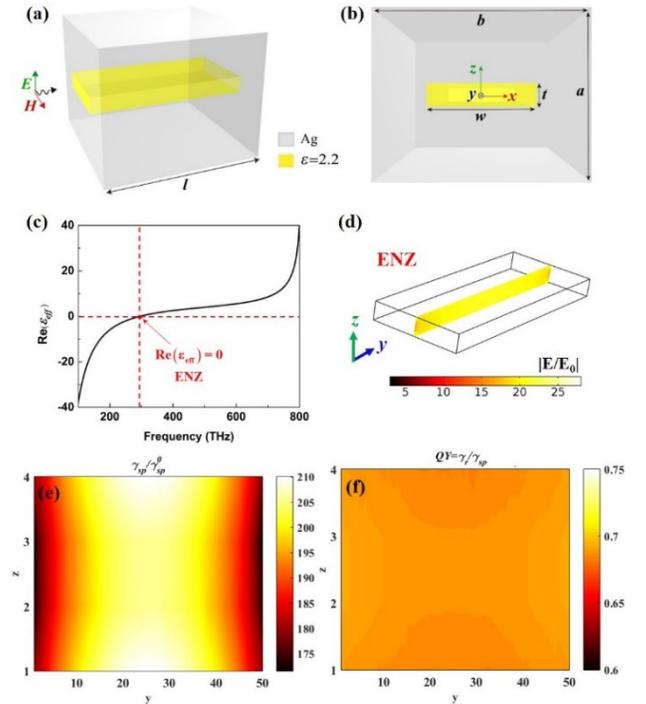

Fig. 6. (a-b) Unit cell geometry of an ENZ orthogonal plasmonic waveguide array. (c) The waveguide's effective permittivity $\epsilon_{eff}$ has a near-zero real part at the cutoff frequency. (d) Electric field enhancement distribution of ENZ plasmonic waveguides at the ENZ frequency. (e) Normalized spontaneous

decay rate $\gamma_{sp}/\gamma_{sp}^0$ and (f) quantum efficiency $QY$ distributions of one emitter placed inside the ENZ plasmonic waveguide. Reproduced from [48].

## B. Superradiance

Superradiance is a fundamental quantum optical phenomenon relevant to the collective photon emission process by many similar quantum optical emitters. This process was first predicted by Robert Dicke [94] in 1954 in the context of collective spontaneous emission and reinforcement of correlations between initially independent atoms or molecules. As shown in Fig. 7(a), in a dilute atomic system, the photon emission by each atom can be considered as an independent spontaneous transition over the characteristic time $\tau_0$. In this case, the emission obeys an exponential decay law and the radiation pattern is essentially omnidirectional. The radiation intensity becomes proportional to the number of atoms $N$. However, these features are notably different when the atomic ensemble becomes dense enough. When the wavelength of light is much greater than the separation of the emitting atoms, as depicted in Fig. 7(b), the atomic ensemble starts to radiate directional photons with much faster and stronger emission compared to independent atoms. This collective emission effect is called "superradiance" [95]. Essentially, this phenomenon is due to the indiscernibility of atoms with respect to the photon emission, which results in constructive interference in the photon emission by the ensemble. The collective radiation of superradiant light behaves as a high intensity pulse with rate proportional to $N^2$ and short emission duration ($\tau_s$) on the order of $\tau_0/N$, which is demonstrated in Fig. 7(b). The counterpart of this enhanced radiation mechanism is called subradiance, a destructive interference process leading to a reduced decay rate from a collection of quantum emitters [96].

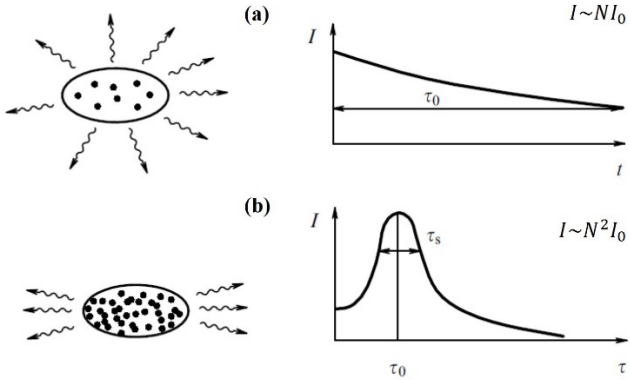

Fig. 7. Comparison between the general characteristics of ordinary spontaneous emission (fluorescence) and superradiance. (a) Ordinary spontaneous emission is essentially omnidirectional in space with exponential decaying intensity. (b) Superradiance is directional in space with an emission occurring in a short burst with duration $\tau_s \sim \tau_0/N$. Reproduced from [95].

Several recent papers exist where researchers propose to achieve superradiance effects and enhance the collective coupling and emission of quantum emitters by using plasmonic nanostructures [36], [48], [97]–[99]. The sub-diffraction confinement associated with plasmonic resonances can be utilized not only to enhance the coupling between a single quantum emitter and the SPP mode, but also to enhance interactions between several quantum emitters, leading to superradiant and subradiant collective emission states [41]. For instance, a collective radiative behavior of $N$ emitters near a metal interface was theoretically analyzed in [98]. It was found that the phenomena of superradiance and surface plasmons can be combined to amplify the emitted radiation intensity $S$ to become $S = N^2 S_0$, which is much higher compared to the single emitter's radiation intensity $S_0$ in free space. Furthermore, many emerging 2D plasmonic waveguide systems were used to study the quantum superradiant effect [36], [47], [48]. Figure 8(a) shows a metallic wedge plasmonic waveguide interacting with two quantum emitters. This interaction includes radiated photons by surface plasmons, and nonradiative excitations (heating) induced in the metal. In this instance, the total decay rate based on the NLDOS can be calculated by Eq. (18). Then a normalized decay rate $\gamma$ is defined by the total decay rate of two interacting emitters divided by the sum of each single emitter decay rate in the same environment, and expressed as following:

$$\gamma = \frac{\gamma_{11} + \gamma_{12} + \gamma_{22} + \gamma_{21}}{\gamma_{11} + \gamma_{22}}. \tag{20}$$

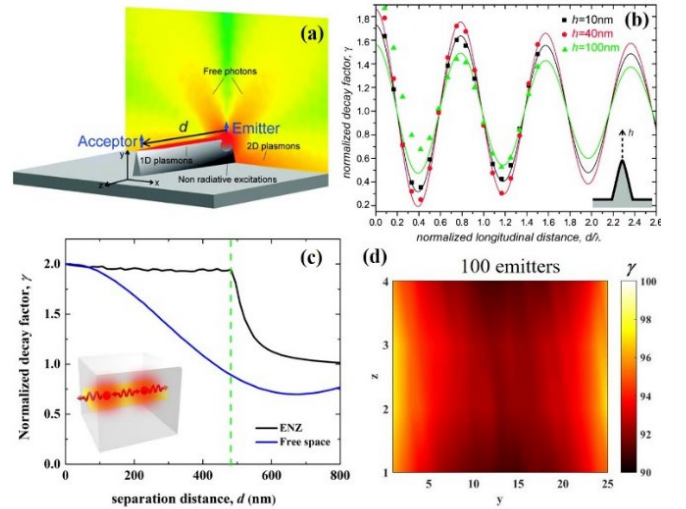

Fig. 8. (a) Schematic and field distribution of a plasmonic wedge waveguide loaded with two quantum emitters. (b) Normalized decay factor $\gamma$ of two identical emitters as a function of inter-emitter separation distance $d$ shown in (a). This plot indicates that both superradiant and subradiant states can be achieved by this system. (c) Normalized decay factor $\gamma$ versus the inter-emitter distance $d$ at the ENZ resonance of the plasmonic waveguide (black line) and free space (blue line). (d) Normalized decay factor distribution ($\gamma$) of 100 emitters uniformly embedded inside the ENZ plasmonic waveguide shown before in Fig. 6(a). Each position on the 4×25 grid corresponds to one emitter. Figures reproduced from: a, b [36]; c, d [48].

Figure 8(b) shows the normalized decay factor $\gamma$ numerically calculated as a function of inter-emitter spacing distance $d$. When $\gamma > 1$ ($\gamma < 1$) the system exhibits superradiant (subradiant) response. It can be seen from Fig. 8(b) that the plasmonic-mediated interaction can efficiently couple to two emitters, enabling the emergence of superradiant and subradiant collective states and leading to substantially modified decay rates [36]. Moreover, recent results demonstrated that both superradiant and subradiant radiation can be realized by using metallic core-shell nanoparticles or plasmonic gratings of paired silver nanostrips coated with dye molecules [99].

However, ideally, this collective coupling will need to be independent of the emitters' locations. A major challenge will be to simultaneously maximize emitters' coupling efficiency into photonic modes and increase their collective spontaneous emission. For example, in the scenario shown in Fig. 8(a), the cooperative behavior of two emitters (superradiance) is very sensitive to their spatial locations in the metallic waveguide system, as it is proven in Fig. 8(b). This directly limits the potential applications of superradiance achieved by plasmonic structures, particularly in the practical scenario of several quantum emitters randomly dispersed along these plasmonic channels.

Recently, it was derived that the superradiance effect can be significantly boosted in ENZ environments compared to conventional materials [47], [100]. This is mainly due to the fact that the ENZ response extends the effective wavelength along the ENZ structure, such that all the emitters feel the same coherent and homogeneous field distribution, as was shown before in Fig. 6(d). For instance, we consider two quantum emitters embedded in the ENZ plasmonic waveguide shown in Fig. 6(a). The normalized total decay factor $\gamma$ is computed by using the Eq. (20) and the corresponding results are plotted in Fig. 8(c) as a function of the inter-emitter distance $d$, where the dashed green line represents the location of the channel's end. Both emitters are assumed to operate at the same frequency close to the ENZ resonance. Interestingly, pure superradiant emission with $\gamma = 2$ is achieved from the emitters inside the waveguide that is independent of the inter-emitter distance along the entire channel. However, the dipole-dipole interactions diminish drastically for emitters placed in free space (blue curve in Fig. 8(c)) and the superradiant effect ($\gamma > 1$) is only achieved when the inter-emitter distance is a few nanometers [101]. More importantly, a collection of multiple identical quantum emitters can be incorporated inside the ENZ waveguide, resulting in a significant boosting of the superradiance emission. Figure 8(d) demonstrates the collective distribution of the normalized decay factor $\gamma$ when 100 identical emitters with the same frequency (ENZ resonance) are embedded in the waveguide. Note that the distribution of $\gamma$ is uniform with values very close to 100 along the entire nanochannel, indicating that all emitters constructively interact with each other leading to superradiant response that is independent to the emitters' locations. Therefore, a random arrangement of quantum emitters will exhibit superradiance when ENZ plasmonic waveguides are used, a very advantageous and unique feature that is expected to be very suitable in the practical implementation of superradiance. In addition, apart from the superradiant emission effect, many other widely investigated quantum optical coherent emitter interaction processes, such as the Förster resonance energy transfer and quantum entanglement will occur, when the distance between quantum emitters is very small [53]. These effects are related to the real part of the Green's function and can be computed by using Eq. (19). Hence, ENZ mediated superradiance promises to have various applications ranging from quantum entanglement to quantum memory and communication systems on a chip. It will also lead to the design of efficient quantum optical memories [102], low-threshold nanolasers [103], coherent thermal sources [104], [105], and ultrasensitive optical sensors [97].

## VI. CONCLUSIONS

In this article, we have reviewed several important aspects relevant to the emerging field of quantum plasmonics [40]. This new research area has attracted and will continue to attract considerable interest among the electromagnetics and quantum optics communities. The existing development of various plasmonic nanostructures, especially ENZ and other plasmonic waveguides, was demonstrated to considerably enhance many fundamental quantum electrodynamic phenomena, mainly based on dipole-dipole interactions. The role of the classical Green's function formalism in quantum electrodynamics has been analyzed and applied to model a diverse range of quantum optical phenomena. The presented plasmonic waveguide designs are expected to lead to a plethora of new nanophotonic quantum optical technologies and devices, such as ultrafast light-emitting diodes, low power plasmonic nanolasers, and single and/or entangled photon sources.